\documentclass[smallextended]{svjour3}
\smartqed

\usepackage{makeidx} % allows for indexgeneration
\usepackage{graphicx}
\usepackage{subfigure}
\usepackage{natbib}
\journalname{Ambient Intelligence and Humanized Computing}

\begin{document}

% Title
\title{Secure System based on UAV and BLE for improving SAR Missions}
\titlerunning{UAV based Rescue System for Emergency Situations}

% Author
\author{Mois\'es Lodeiro-Santiago \and Iv\'an Santos-Gonz\'alez \and Pino Caballero-Gil \and C\'andido Caballero-Gil}
\authorrunning{Mois\'es et al.}
\institute{Departamento de Ingenier\'ia Inform\'atica y de Sistemas, \\Universidad de La Laguna, Tenerife. Espa\~na\\
\email{\{mlodeirs, jsantosg, pcaballe, ccabgil\}@ull.edu.es}
}

% Received and Accepted
\date{Received: date / Accepted: date}

% --------------------------------------------------------

\maketitle% typeset the title of the contribution
%19.144

\begin{abstract}
This work describes an integrated solution to face a civil security problem in the area of Search And Rescue (SAR) of missing people. This proposal is based on the use of emerging technologies such as Unmanned Aerial Vehicles (UAV), also known as drones, and the use of simulated beacons on smartphones. In particular, in the presented tool, drones fly synchronously in a specific area so that each drone uses on-board sensors to scan and detect any signal emitted by Bluetooth Low Energy (BLE) beacons from smartphones of missing people. This technique allows getting the GPS position of any detected missing person. This work also includes some security issues related to possible attacks focused on the perimeter and physical security.
\end{abstract}

\section{Introduction}

In the last years, outdoor sports activities have become one of the most favourite pastimes. Mountains are usually the most popular place for practising outdoor sports in spite of the fact that these activities usually involve going to dangerous places where people can have to face critical situations. Indeed, mountains are one of the places where more Search And Rescue (SAR) operations are undertaken.
Most situations where a SAR team is needed are usually caused either by bad weather, or negligence, carelessness or lack of concentration of people. 

The typical procedure when someone gets lost is as follows. A SAR team is deployed to the area where the person might be. The timeout period elapsed prior to completion of the SAR operation can make rescue tasks more difficult due to fainting, anxiety, sunstroke, etc. When a SAR mission starts, nobody knows how long it will take because it depends on the number of police officers, fire-fighters or civilians involved in the operation. Time is one of the main problems of SAR missions because when someone gets lost, in the middle of a mountain, for example, the period till being rescued could determine the health and even the life of the missing person. Sometimes, helicopters are needed in SAR operations to facilitate search work, but rarely more than one can be used due to the high costs involved in each SAR operation.
	
In this work, we have developed a novel solution to face the difficulties involved in each SAR operation. In particular, this proposal consists of a platform to synchronize SAR operations involving the use of drones to perform the search tasks that would nowadays be performed by helicopters in combination with people, but without the high costs that this would involve. The way a person will be located within this proposal is by making use of signals emitted by beacon devices using Bluetooth Low Energy. These signals can come from mobile devices such as smartphones, or Bluetooth gadgets such as Physical Web beacon devices developed by Google.

This work is structured as follows.
Section \ref{sec:preliminaries} describes the used beacon technology and introduces some basic concepts on Unmanned Aerial Vehicles. Section \ref{sec:rescuesystem} presents the state of the art on technological solutions used for SAR operations. Section \ref{sec:beaconimpl} includes the main details of beacon signalling implementation in smartphones. Section \ref{sec:emergencyprotocol} describes in detail the steps followed in the proposal to successfully complete a SAR mission. Section \ref{sec:security} contains a study of some possible attacks that can affect the proposal. Section \ref{sec:benchmark} shows a comparative study between the current searching methods and this proposal. Finally, Section \ref{sec:conclusions} closes the paper with some conclusions and open problems.

\section{Preliminaries}
\label{sec:preliminaries}

Two different technologies are combined in the proposal presented in this paper. The first one is beacon technology, which is a low energy device that emits a broadcast signal thanks to Bluetooth Low Energy (BLE) technology. Traditionally, beacons have been used in indoor positioning systems \citep{inoue2009indoor} \citep{yim2008introducing}, for which they obtain accurate locations thanks to the short range that beacons offer. Besides, beacons have been also widely used in the commerce world where for instance when a client is approaching a certain shop or product, he or she automatically receives an advertisement beacon signal with discounts for that specific shop/product.
 
Beacon performance depends on the specific hardware, which is usually composed of a BLE chip and a power battery. The use of BLE technology allows beacons to be used with minimum power consumption, so the battery of a beacon usually resists for a long time, and even for years \citep{kamath2010measuring}.
A beacon device usually broadcasts its Universally Unique IDentifier, which uniquely identifies the beacon and the application that uses it. In addition, the broadcast signal also includes some bytes that can be used to define the device position, send a URL, or automatically execute an action.
 
In particular, the performance of beacon devices is based on two main properties of beacons: the Transmission Power (TX Power) and the Advertising Interval. The TX Power is the power with which the beacon signal is broadcasted. This transmission power can be manually established. On the one hand, a bigger TX Power implies that the beacon can be discovered in a larger distance due to the properties of the signal intensity (that decreases with the distance), but implies bigger battery consumption too. On the other hand, a lower TX Power implies a minor range but also a lower battery consumption. The Advertising Interval is the frequency with which a beacon signal is emitted. This frequency can be established, depending on the necessity of the developed system and its latency requisites. The Advertising Interval is established in milliseconds so that bigger values imply bigger battery consumptions, while lower values involve lower battery consumptions.
Different protocols can be used to send data in packets through the Advertising mode. These protocols establish a format in the advertising data so that the applications can read these data in a proper way. The most extended protocols are: iBeacon \citep{ibeacon}, created by Apple; Eddystone \citep{eddystone}, which is the open source project created by Google; and AltBeacon \citep{helms2015altbeacon}\citep{github:altbeacon}, which is the protocol developed by Radius Networks. 

The second technology used in the proposed system corresponds to Unmanned Aerial Vehicles (UAVs), which are unmanned aircraft vehicles controlled either remotely by pilots or autonomously by on-board computers with GPS technology. There are two different kinds of UAVs: gliders and multi-copters. On the one hand, gliders are UAVs that, as their name indicates, can glide, what is an important advantage over other type of drones because it supposes a relevant improvement on the energy consumption. On the other hand, multi-copters are UAVs that have more than one helix. Their main advantage is the ability to be static in the air at the expense of higher power consumption. 

Autonomous UAVs are the types of UAVs that are more interesting for the proposal. In this regard, there are mainly two different systems that have been used to get an autonomous operation. The first one is known as Inertial Navigation System (INS) \citep{berg1970inertial} and uses the output of the inertial sensors of the UAV to estimate its position, speed and orientation. In these positioning systems, the UAV uses a 3-axis accelerometer that measures the specific force acting on the platform and a 3-axis gyroscope that measures its rotation. An advantage of this kind of positioning systems is that it does not require the use of any other external signal to provide an external solution. On the other hand, this type of systems produces positioning errors of the order of several hundreds of meters. The other main kind of positioning system used by the UAVs is the GPS system. Since the GPS system needs an external signal to work, and the navigation accuracy depends on the signal quality and the geometry of the satellite in view, depending on the used GPS, the error can vary between one and hundreds of meters. To solve the problems of these navigation systems in UAVs, it is usual to combine INS and GPS to get a good precision from the GPS system and the frequency positioning data of the INS. This solution is also applied in the proposal here described.

A mandatory prerequisite for the proper operation of the system is that the devices (smartphone and drone) have enough charge to work throughout the rescue mission. If the drone (or smartphone) does not have enough battery, this could produce money and information losses (mainly because of the hardware). In spite of the fact that this paper does not take into consideration other supply power system, it could be possible to feed the air vehicles as in other aircraft such as the "Solar Impulse 1" or as it is proposed in the patent of the Boeing company \citep{boeingPatent} where a recharging system is defined by a static flight over a series of charging-antennas. However, this kind of improvements would not allow current drones to fly due to the weight, surface area, load balancing or hardware cost.

\section{State of the Art}
\label{sec:rescuesystem}
During the last years, different proposals have been presented in the field of SAR systems. Some of them are only communication systems, but others include different rescue mechanisms. A lot of them focus on the use of the so-called emergency call, included nowadays in all mobile devices. This mode allows users to make emergency calls, even when they have no coverage to make a phone call to any of their contacts. 
To make this system work, the European Union decided in 1991 that its member states had to use the 112 number for emergency calls. Moreover, mobile operators and mobile phone manufacturers were forced to adapt all their products to let users operate with the network infrastructure of any other operator to make emergency calls, independently of the mobile operator. This ability is the basis of Alpify \citep{alpify}, a mobile application that allows using the emergency call system when there is no Internet connection, and notifying emergency services and contacts through SMS including the georeferenced coordinates of current position. A problem of this proposal that derives from the use of the emergency call system, is the necessity of availability of some network infrastructure because it requires coverage of at least of one operator. 

In the situation analysed in this work, the initial hypothesis is that there is no coverage of any mobile operator because all network infrastructures are unusable, collapsed, or simply do not exist because the place is remote or difficult to access. In these cases, the use of a system based on communications through network infrastructures is not a feasible solution. Moreover, the use of systems that need previous agreements between the company that develops the system and the different emergency services of every state of each country, makes it more difficult to use them in different places around the world. 

Centralized systems like Alpify only notifies the emergency services and the contact person previously selected, and depending on the location, they may be delayed in arriving to the position. The fact of not letting people who are near help in a faster way, supposes an important disadvantage of this kind of systems. 

Other systems, like the one described in \citep{santos2014alternative}, propose the use of Wi-Fi Direct technology, so in the case where there is no Internet connection, people that are in the range of the Wi-Fi Direct technology are notified. The basis of this system consists of sending an emergency message that includes the geolocated position of the person, to other people in the range of Wi-Fi Direct. When this emergency message is received, if the user has Internet connection, he/she automatically notifies the warning to the emergency services. If the user does not have Internet connection, the message is re-sent through broadcast in a multi-hop system. This system involves an improvement compared to other systems than only use the emergency call, but it has problems when the number of users is not enough or they do not have Wi-Fi Direct technology in their smartphones. 

The authors of the work \citep{jang2009rescue} try to solve this problem through the use of other devices like laptops carried by the emergency services. In that proposal, the goal is that the first units of the emergency services that arrive to the place deploy a Mobile Ad-hoc Network based on Wi-Fi through laptops. After the network deployment, a group communication system based on peer-to-peer architecture, which admits communications such as VoIP, Push-to-Talk, instant messaging, social networks, etc. is proposed. The main advantage of this kind of solutions is that the network is managed by specialized personnel and the use is restricted only to emergency services, preventing the access of other users to the network. Its main disadvantage is that people who have suffered a catastrophe and are isolated, continue being isolated, during the time that the emergency services take to arrive to the catastrophe place. 

Different schemes, like the one described in \citep{lamarca2005place}, use Physical Web beacons to find injured people, once the emergency or rescue protocol is activated. This kind of systems has the disadvantage that the search has to be done manually, and the rescue or emergency service must have access to the emergency zone, which might not be possible in some situations. 

Other proposals use UAVs for SAR operations in emergency situations where rescue teams cannot access. This is the case described in \citep{dickensheets2015avadrone}, where the use of UAVs in combination with Physical Web beacons is proposed to search people affected by avalanches. 

Moreover, there are systems, like the one described in \citep{ezequiel2014uav}, which use aerial imaging processing to detect people in problems for post-disaster assessment. Their main disadvantage is that the use of image processing could be imprecise in situations where the visibility is not good, as can be in fires where smoke hinders visibility. 

Other solutions, like the one proposed in \citep{chou2010disaster}, for the management and monitoring in emergency situations are based on the use of real-time aerial photos. The main advantage of that proposal is the use of real-time aerial photos to produce a digital elevation model data that can be very useful to see the latest terrain environment and provide reference for disaster recovery in the future. 

The work \citep{Silvagni201718} offers solution related with the proposal of this paper, which is composed of a body detection system using sensors on drones in snow avalanches (and ski areas), forest and mountains. The authors make use of two cameras: the normal day camera and one thermal for detection in programmed night flights. They also discuss the possibility of including a weight of up to 5kg as a payload package. Another alternative is the proposal of \citep{bejiga2016convolutional}, which describes the use of UAVs equipped with cameras to find people missing in snow avalanches using neural networks trained to detect changes in patterns. The proposal here described be complementary to both proposals in order to improve them with remote administration and including real-time monitoring.

\section{Beacon Implementation}
\label{sec:beaconimpl}

The use of beacons implies an important advantage in SAR operations thanks to the higher speed and accuracy of operations for locating missing people who carry beacons. However, a disadvantage of this system is the need to carry a device with us to allow being rescued if an emergency situation happens. That is the reason why in this proposal smartphones are used, because they are devices that people always carry with them.

In order to use the beacon system in the mobile platform, the best solution is the use of BLE wireless technology to simulate the beacon. In the Android operating system, this is possible in all the devices with 5.0 and higher Android versions and Bluetooth 4.1 compatible devices. To implement the simulated beacon, the Android BLE peripheral mode is used to allow running applications that detect the presence of other smartphones nearby, with minimum battery consumption. 

For the implementation the AltBeacon Android Beacon Library \citep{github:altbeacon} has been used. This is an open source project that allows detecting beacons, meeting the open AltBeacon standard by default. Besides, it can be easily configured to work with the most popular beacon types like Eddystone or iBeacon formats. In the proposed system, beacons have been implemented using the Eddystone-URL format, a new variant of Google UriBeacon, also known as the Physical Web beacon. The main difference of the Eddystone-URL format is that it does not transmit an UID but a URL that can be automatically detected by the smartphone browser or an application and open the URL in the browser without user intervention. This system uses 17 bytes to store the URL, fact that makes the use of short URLs necessary. However, this is not a problem because there are a lot of services, including one developed by Google, which allows shortened URL.

\section{Proposed System}
\label{sec:emergencyprotocol}

In order to design a system that can provide a complete solution on this field, the described proposal has been developed in two different parts. On the one hand, an active search solution is offered combining emerging technologies like UAVs and BLE beacons. On the other hand, Bluetooth and the help of citizens are combined to locate missing people. These two parts of the system are below named active and passive method, respectively.

\subsection{Active method}

The so-called active method is focused on the use of Bluetooth protocol devices that emit a spherical radio signal containing a unique identifier for each device so that when a user gets lost, he or she could be located by performing a scanning of Bluetooth signal in order to determine his/her approximate position. This scan would be performed over the area using one or more drones.

First of all, every user must have an Android device and register his/her user data, such as name, surname, address, blood type, etc., using the system web platform, which must be handled by the operator of the emergency service. Then, the operator, once the user is registered, provides the code that the user must introduce in his/her terminal to be paired with the database user profile. Afterwards, if a user gets lost, the SAR steps would be as shown in Fig.

\ref{fig:nuevoEsquema}:
\begin{figure}[ht]
	\centering
	\includegraphics[width=1.35\textwidth]{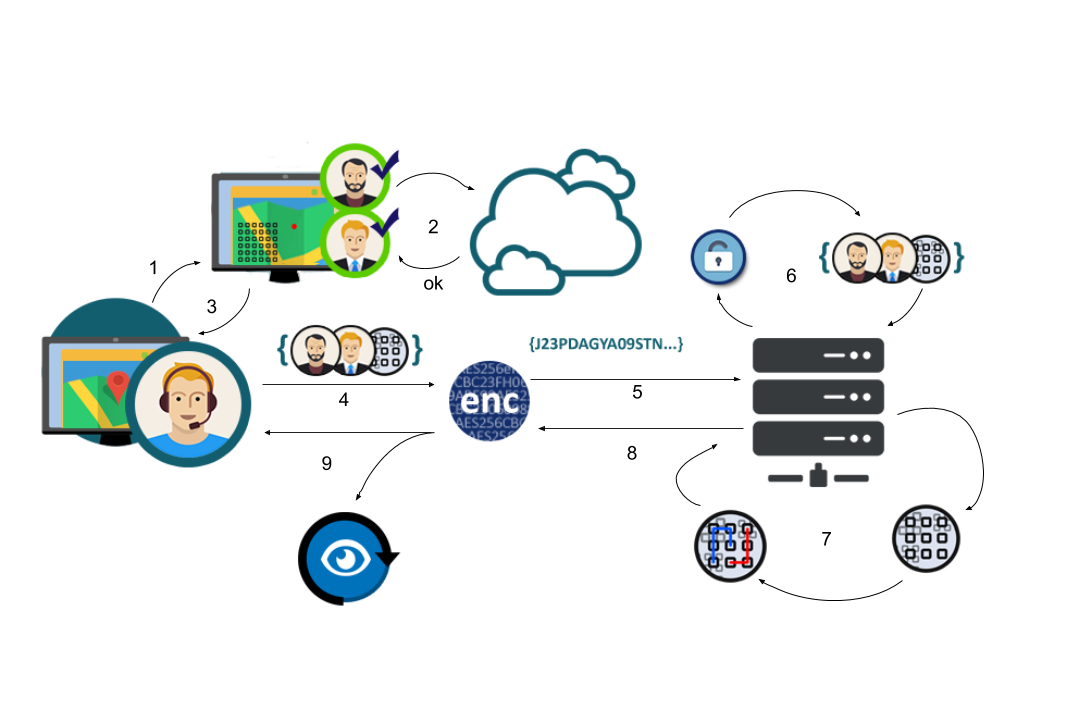}
	\caption{Flow diagram of the proposed emergency protocol}
	\label{fig:nuevoEsquema}
\end{figure}

\begin{figure}[ht]
	\centering
	\includegraphics[width=1\textwidth]{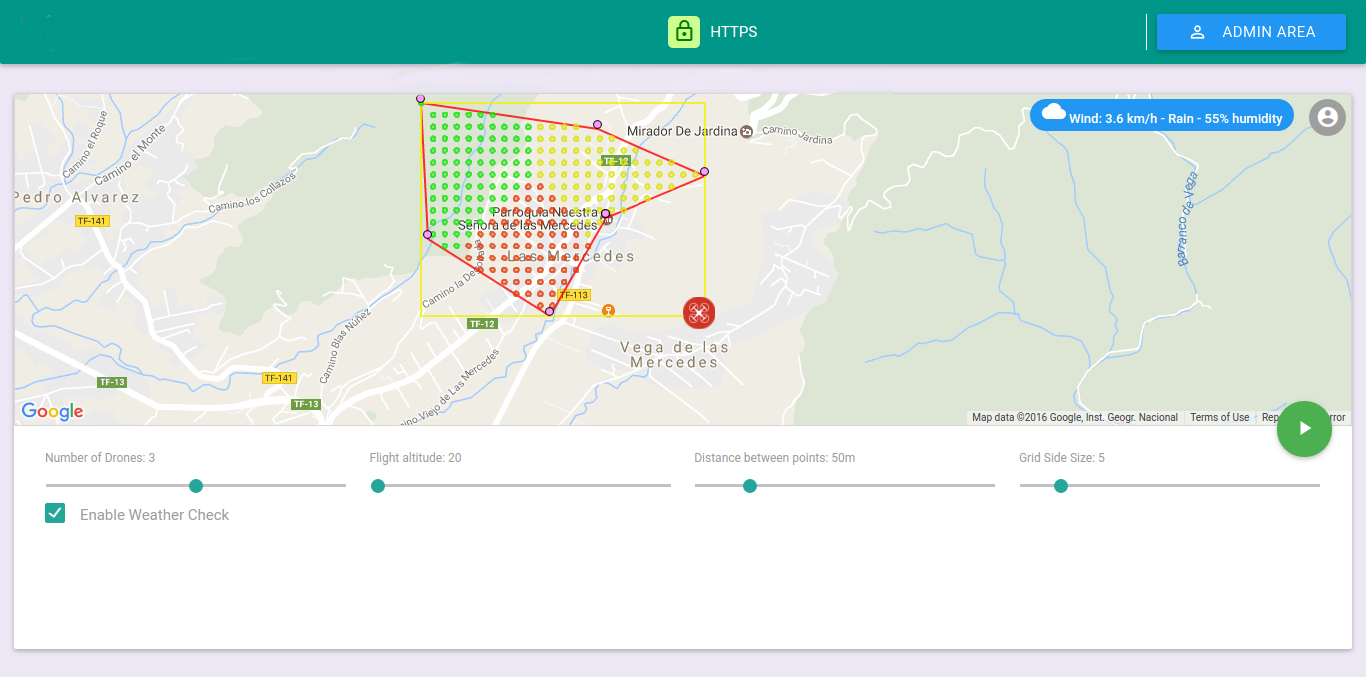}
	\caption{Operator web panel}
	\label{fig:nuevoPanel}
\end{figure}

\begin{figure}[ht]
	\centering
	\includegraphics[width=1\textwidth]{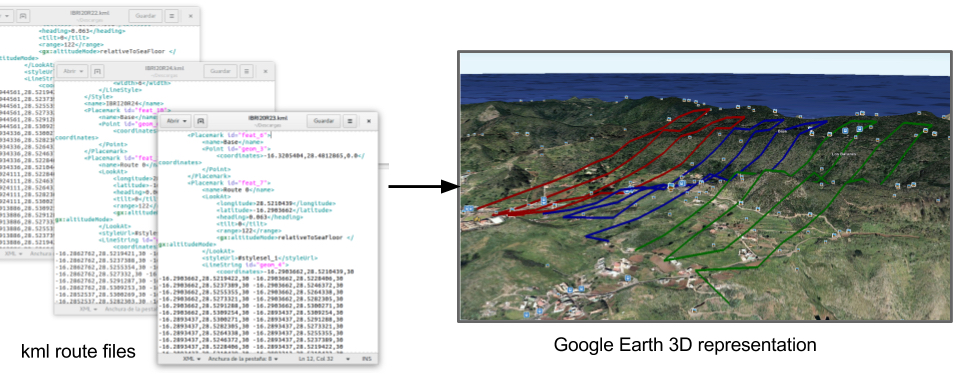}
	\caption{Generated KML and view in 3D}
	\label{fig:kml3d}
\end{figure}

\begin{figure}[ht]
	\centering
	\includegraphics[width=1\textwidth]{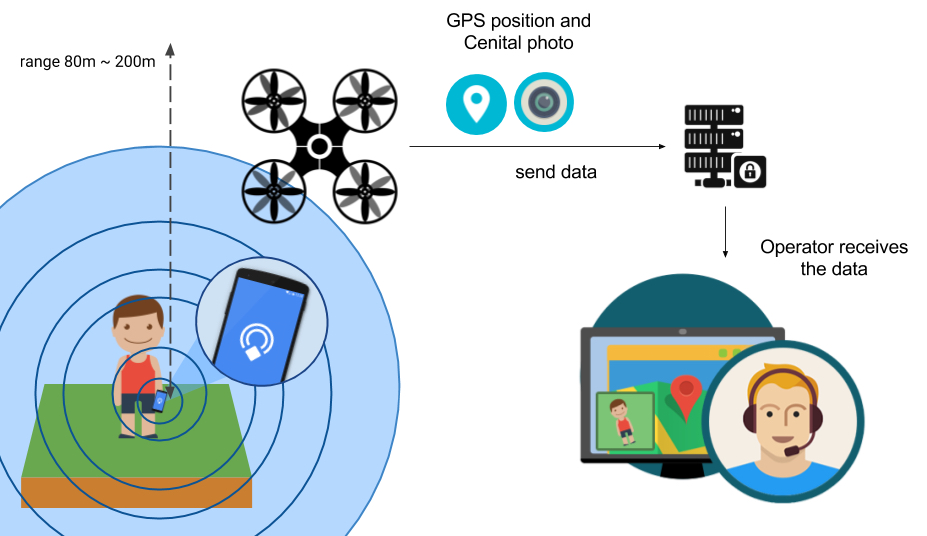}
	\caption{Drone detecting a missing user}
	\label{fig:dronedetection}
\end{figure}

\begin{itemize}
	\item \textbf{Step 1}: The operator, through a phone call, is notified by a contact person about the disappearance of a registered person. Then, the operator launches a control panel web platform (shown in Fig. \ref{fig:nuevoPanel}). This panel shows in its main screen an interactive Google Map that allows the operator to intersect with the back-end system through a few clicks. For instance, using it, the operator can click on the map to set a series of points in order to create a search area. The panel also allows selecting several parameters such as the user/s in search, the number of drones in the mission, relative flight altitude with respect to the floor and distance between points inside the polygon.
  
	\item \textbf{Steps 2 and 3}: When the operator completes the mission configuration, he/she clicks on the 'Start Button' that triggers a signal to start the SAR process. Then, the system automatically checks the climatology in that area in order to verify the weather and decide whether to launch the mission or not. If the weather is good enough, the mission continues. Otherwise, if the weather is, for example, rainy or windy, the operation based on drones gets cancelled because drones cannot fly in such weather conditions.
	The panel also includes a check-box to continue the mission despite the bad weather.
	
	\item \textbf{Steps 4 and 5}: Data are then sent using the encryption algorithm discussed in Section \ref{sec:security}, to the main server, packed in an encrypted JSON format, through a secure HTTPs layer.
	
	\item \textbf{Steps 6 and 7}: The server decrypts the received data in order to get the identifiers of user(s) in search, and the map grid points used to calculate the grids of each searching area. Using the generated grid points, and starting from the established base point (red dot in Fig. \ref{fig:nuevoPanel}), an optimal route is calculated for each area. The way to make sure that all points are visited is by applying a solution to the Travelling Salesman Problem (TSP) in order to provide a low-cost route. Thus, the path-finding algorithm A* is used in each area as a point graph \citep{astar}. All areas take the same starting point, so to solve and obtain separate routes, the server processes each area separately, and a Keyhole Markup Language (KML) is used to display geographic data in three dimensions (see Fig. \ref{fig:kml3d}).
	
	\item \textbf{Steps 8 and 9}: These steps run in parallel, being the most important ones in the emergency protocol because without them, the use of drones in the mission has no sense. When the drones take off (each one to its search area), the mission starts. Every few seconds, the web operator panel asks to the server asynchronously whether it has new updated data. These updates are sent by the drones constantly. If the smartphone device, which goes on top of the drone, finds a Bluetooth signal from a beacon in search, it checks with the main server whether that person is being searched or not. If so, the drone's smartphone automatically begins to take several cenital photos that the operator can see in real time from the web panel as if it were a streaming service (see Fig. \ref{fig:dronedetection}).
\end{itemize}

\begin{figure}[ht]
	\centering
	\includegraphics[width=1\textwidth]{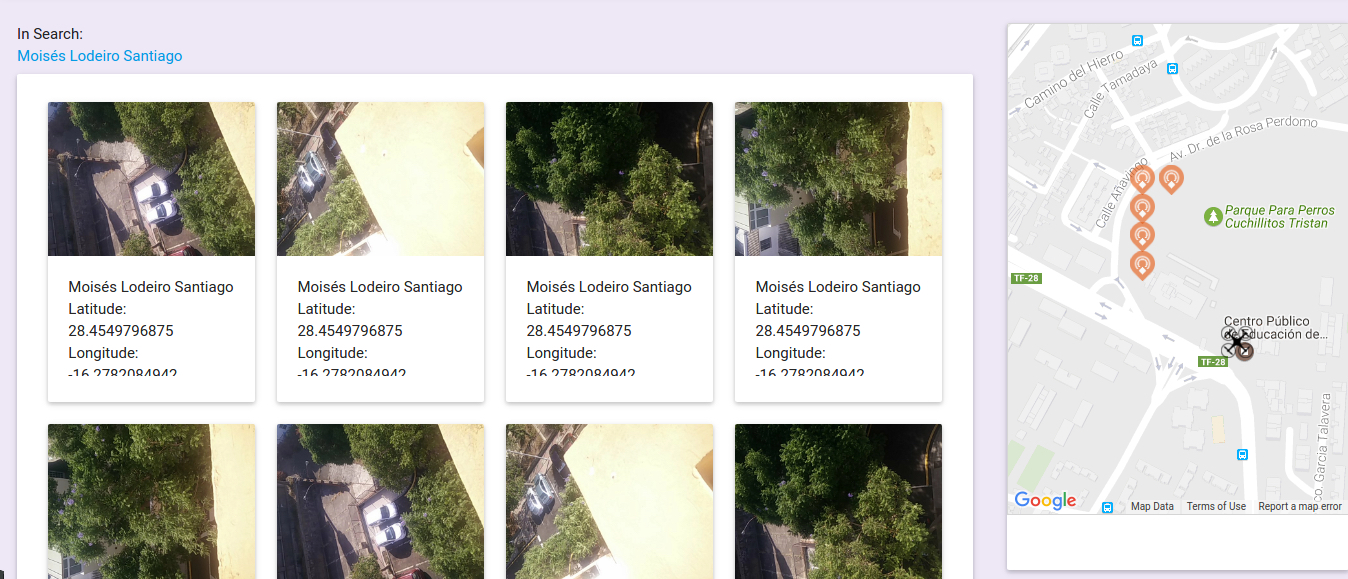}
	\caption{Result panel}
	\label{fig:resultados}
\end{figure}

Once the drone has finished the flight, the operator can analyse all the information gathered by the smartphone
and view the results of past missions on the operator panel (see Fig. \ref{fig:resultados}). In this panel, the operators can see all the photos that the drones have taken and the GPS positions for each one. The rescue tasks begin when the lost person has been found.

\subsection{Passive method}

\begin{figure}[ht]
	\centering
	\includegraphics[width=1.35\textwidth]{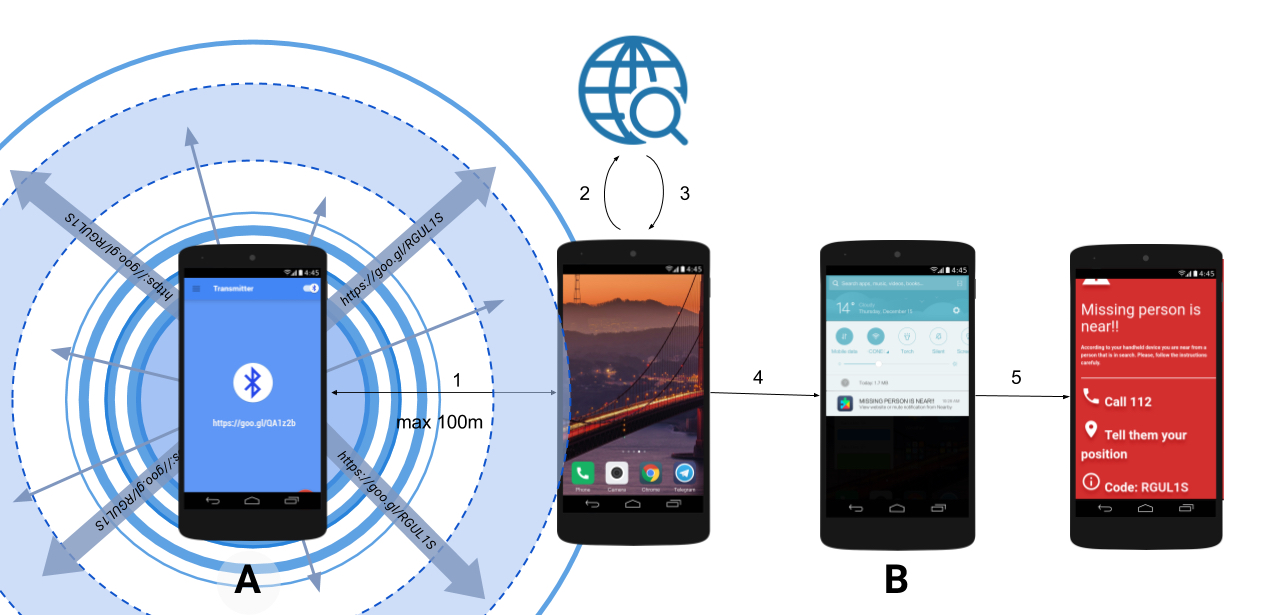}
	\caption{Passive detection protocol}
	\label{fig:passiveMethod}
\end{figure}

One of the main advantages of using BLE signals, especially those that apply to the UriBeacon protocol or the new Eddystone, is the ability to store a web address in the physical device or in the simulated beacon. The URL used in this proposal is a web address that points directly to a user profile. The meaningful question here is that, with no need to install a new application on the mobile phone, the smartphone is able to display beacon notifications with the received BLE signal information. This notification reflects the header title from the website that the sender is emitting. The detailed steps are explained in the list shown below. This is possible thanks to the fact that in Android 5.1 or higher, Google has implemented a system within its 'Google Now' system that allows passive scanning of beacon devices nearby, following five steps:

\begin{itemize}
	\item \textbf{Step 1:} The simulated beacon (see Fig. \ref{fig:passiveMethod}-A) sends a 100 m wireless spherical BLE signal that contains a URL paired with the user profile
	\item \textbf{Step 2:} If there is a smartphone (see Fig. \ref{fig:passiveMethod}-B) inside the emitted signal range, the smartphone takes the received shorten URL and visits it in the background.
	\item \textbf{Step 3:} The server sends the HTML header (title, body and related information) to the smartphone.
	\item \textbf{Step 4:} The device shows it as a notification as seen in Fig. \ref{fig:passiveMethod}-B.
	\item \textbf{Step 5:} When the user clicks on the alert, a web browser is opened to display the content of the page. In this case the displayed data correspond to three steps:
	\begin{enumerate}
		\item Call 112 - It is necessary that the user can rely on the system when he or she makes the emergency call through the 112 phone number.
		\item Communicate position - This step allows the emergency squad to determine where the user is.
		\item Code - The user must introduce the code that appears in the screen, corresponding to its identifier.
	\end{enumerate}
  
\end{itemize}

The power of the feature mentioned in the last step has been used to make people can participate in the search for missing people since, when a person disappears, anyone with an Android terminal with the aforementioned version could be very useful for the SAR operation by using its smartphone in collaboration with the emergency staff. If the person detects a BLE signal that corresponds to a beacon, but the user is not in search, the emitted URL will point to a 404 page (not found) and it will not be displayed on the smartphone as a notification. This is performed dynamically and without any user interaction.

\subsection{Route planner}

It is estimated that the average maximum flight time in a drone is now half an hour or a little more. That is the main reason why saving time in search is a fundamental problem in this proposal. In a first approach, the search areas were established as equidistant geographical points making use of the formula of the semiversene (shown below) where $\varphi$ is the latitude, $\lambda$ is the longitude and $R$ is the earth radius (6.371 km approximately). In order to use this formula later to define a square cloud of points that drones have to visit in an orderly manner, each one starts from a point on the edge of the initial dotted square.
\\\\
{
\label{eq:haversine}
$a = sin^2\frac{\Delta \varphi}{2}+cos\varphi_{1}*cos\varphi_{2}*sin^2\frac{\Delta \lambda}{2}\\
c = 2 * tan^{-1} \frac{\sqrt{1-a}}{\sqrt{a}}\\
d = R*c$
}

The main problem of this solution is that on many occasions the square areas are too large while the main area of search is small compared to the other areas. That is why we chose to improve the functionality of search in different zones in order to be able to implement the use of irregular polygons. With the implementation of an irregular polygon, another problem arose. How could we separate the areas in an approximately equal way? After investigating different techniques we opted for a method of separation of areas based on the 'k-means' clustering technique.

\subsection{K-means}

The k-means clustering algorithm classifies $n$ points within $k$ clusters by assigning each point to a cluster where the value of the average in a set of $p$ variables is calculated by some approximation technique such as the Euclidean distance. The algorithm computes these assignments iteratively until, with the reassignment of points and recalculating the distances of cluster points produces no changes. The calculation of clusters using k-means involves identifying a series of points called centroids. There are several methods for finding a set of centroids that gives good results, depending on the data sets. The easiest way is to take a set of random points depending on the space (where the number of centroids depends on the desired cluster's numbers). Then, each centroid has to be assigned to its nearest division. After the assignment update, the centre is added in the coordinates of a new point. Assigning all the points to a particular set and successively updating the centres. When there is no change in centroids, the k-means algorithms stops.
The pseudocode of the algorithm can be seen below:

\begin{itemize}
	\item Input: K, set of points $x_{1}...x_{n}$
	\item Place centroids $c_{1} ... c_{k}$ at random positions
	\item Repeat until convergence:
	\begin{itemize}
		\item for each point $x_{i} \in K$:
		\begin{itemize}
			\item find the nearest centroid $c_{j}$ using the same distance approach between the point $x_{i}$ and the cluster center $c_{j}$
			\item assign the point $x_{j}$ to the cluster of the nearest centroid $j$
		\end{itemize}
		\item for each cluster $j = 1..K$
		\item $\forall a \in 1..d / c_{j}(a)=\frac{1}{n_{j}}\sum_{x_{i} \rightarrow c_{j}}^{}x_{i}(a)$
		\item new centroid $c_{j}$ mean of all points $x_{i}$ assigned to cluster $j$ in previous step
	\end{itemize}
	\item Stop when none of the cluster assignments change
\end{itemize}

\begin{figure}[ht]
	\centering
	\includegraphics[width=1\textwidth]{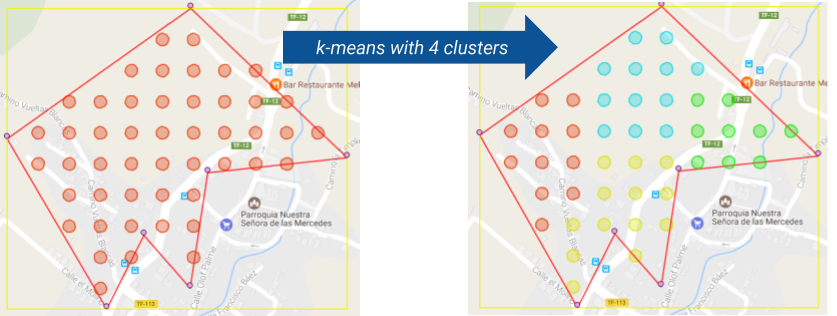}
	\caption{Flow diagram of the emergency protocol}
	\label{fig:kmeans}
\end{figure}

In this proposal, the k-means algorithm is used for dividing areas within an irregular polygon. This means that, having an irregular polygon, we can paint geographically the interior points over a map maintaining the calculated distances D between each pair of points. Once we have the point cloud, we apply the k-means algorithm on it, thus getting N distinct areas (see Fig. \ref{fig:kmeans}). Then, for each of these areas and starting from an initial point for all, the solution is applied to the above mentioned TSP problem by applying A* path-finding algorithm.

\section{Security of the proposal}
\label{sec:security}

First, drones can be targets that are easy to reach, through physical or electromagnetic attacks.

\begin{itemize}
	\item Physical attacks:
	\begin{itemize}
		\item Net: If the drone is reached by a net or any type of clothes, the threads can roll up into the helices and the drone could fall to the floor. 
		\item Metallic dust: If an attacker throws any kind of metallic dust to a drone, the particles of dust can stick into the main motor bringing the drone down.
	\end{itemize}
	\item Electromagnetic methods:
	\begin{itemize}
		\item Signal Jammer: A simple powerful signal jammer can nullify or create jamming into the phone signal or the phone sensors like Bluetooth, Wi-Fi, etc.
		\item GPS Jamming: An attacker can send a fake GPS signal to change the drone route. This is possible when the attacker broadcasts interferences between the drone and the pilot so the estimated drone's location is wrong. Many security strategies have been proposed against jamming in wireless communications.
		\item Data interception: When a drone is in flight and emits a signal through Bluetooth or Wi-Fi, the information travels through the air as a sphere because the signal is emitted in all directions. Thus, anybody can intercept that signal using a satellite dish to read the raw data that the drone is emitting. For this reason, in this proposal sent data are encrypted with a strong algorithm.
	\end{itemize}
\end{itemize}

Secondly, different types of attack on the server can be launched:

\begin{itemize}
	\item Brute force: A brute-force attack is a type of attack that is usually applied to gain access to an account in order to retrieve sensitive data. The way to perform it is by sending many packages, trying tons of possible combinations of usernames and passwords. Filtering the traffic could be a great solution to avoid this problem. Also, disabling all the traffic from the outside to the server could be useful.
	\item Software-based attacks: Malware, Trojans, key-loggers, virus, ransomware, etc. All those techniques are software-based and are normally used to control the machine where they are executed, or in the worst case, to control the drones. This could cause a sensitive data loss in our server. To face that problem, it is necessary the use of an integrated server-protection like a firewall, anti-virus, etc.
	\item DoS/DDoS: A Denial of Service (DoS) is a type of attack that tries to flood traffic on a server machine. When the server is collapsed, it cannot answer all the requests from other users. This makes that the server becomes unavailable for the rest of users. If this attack is launched by many attackers at the same time, it is considered a DDoS (Distributed Denial of Service) attack. In order to avoid this kind of attacks, we use a traffic filter.
\end{itemize}

In order to protect data, some of the implemented data protection measures are as follows.
Database core uses an SQL-based database. Instead of querying over it directly, we use another abstract layer over the database motor, known as Object-Relational Mapping (ORM). This layer allows a programmer to interact with the models of the database as if they were objects from a class. Also, this ORM is improved with a middleware that allows the raw stored data to be encrypted using an AES-256 CBC algorithm. With this method we avoid possible database intrusion using SQL injections. The AES-256 CBC is also the encryption system that is used in each package that the drone sends to the server, and on the BLE communication between the drone and the beacon.

\section{Benchmark}
\label{sec:benchmark}
In this section some questions like reliability of the system, monetary costs and time are responded.

\subsection{Detection Range }
\label{rangedetection}
Physical Web Beacons used in this proposal must be able to be detected from long distances in order to perform the missions correctly. However, to make this possible it is mandatory that the flight altitude corresponds to the emission range of a Bluetooth sensor. Besides, it is necessary that the flight altitude allows avoiding certain obstacles like trees, antennas, etc. 

\begin{figure}[ht]
	\centering  %%% not \center
	\subfigure[Detection range tests in horizontal]{\label{fig:btrange}\includegraphics[width=0.8\linewidth]{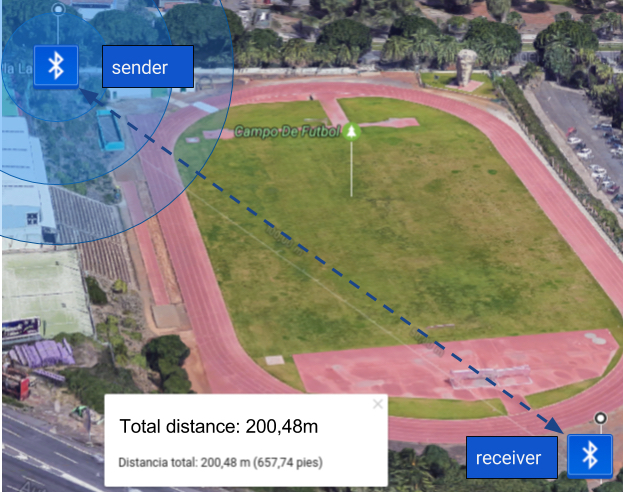}}
	\subfigure[Detection range tests in vertical]{\label{fig:btrange2}\includegraphics[width=0.8\linewidth]{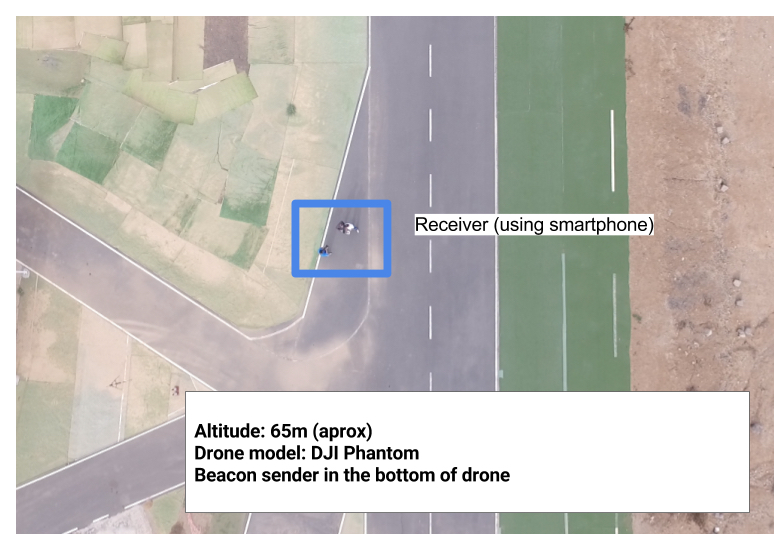}}
	\caption{Bluetooth range detection in horizontal and vertical}
\end{figure}

According to the Bluetooth SIG \cite{btrange}, a BLE range could reach 500 meters in open field.
In order to verify the system's effectiveness, we have carried out an empirical range detection. The used devices have been two smartphones (both with Android OS), one with Bluetooth 4.1 as beacon transmitter and the other one with Bluetooth 4.0 as receiver. The used broadcast application was BeaconToy \citep{beacontoy}. In addition, for the reception we used the Beacon Tools \citep{beacontools}, application made by Google. The battery test was made in a horizontal open field in a day with some scattered clouds, no wind and no obstacles between emitter and receptor (see Fig. \ref{fig:btrange} and \ref{fig:btrange2}). To perform the vertical test, we placed a small Physical Web Beacon in the drone's bottom. Meanwhile, in the ground we were measuring and checking detection range carrying out the same tests done in horizontal test (10 measures for each altitude). In a real environment, the system would be set in reverse order but we assumed that this factor would not interfere in the experiment. For each distance, each test was repeated 10 times, and the results are shown in the Table \ref{tab:detectiontable}.

\begin{table}[htbp]
	\centering
	\caption{Bluetooth signal detection}
	\label{tab:detectiontable}
	\begin{tabular}{|c|c|c|}
		\hline
		Distance & Success Rate \\ \hline
		10\ m   & 100\% \\ \hline
		20\ m    & 100\% \\ \hline
		50\ m   & 100\% \\ \hline
		100\ m   & 100\% \\ \hline
		150\ m    & 90\% \\ \hline
		200\ m  & 60\% \\ \hline
	\end{tabular}
\end{table}

Some studies \citep{bluetoothrange} indicate that the emission range is improving with each new version of BLE technology.

\subsection{Detection time}

The emission power of a Bluetooth device depends on the broadcast period, so that the faster the device emits, the less emission range it has. To perform a search benchmark, a system called Wartes \citep{generalitat} to calculate the probabilities of detecting clues during the search. With this method, the factor $P $ (probability to detect the victim) can be calculated through the equation $Pa * Pd = Pe$ where:

\begin{itemize}
	\item $Pa$ is the value that gives priority to the segment of the search area.
	\item $Pd$ is the value given to the ability of the resources to detect clues of the victim.
	\item $Pe$ is the result of the two previous factors, used as a measure of success.
\end{itemize}

 Table \ref{timeGeneralitat}) shows the needs for a search in an area of 2.5 square kilometres. Using the same distance to cover, with a constant speed of 5 m/s, using drones the result in search time is the one shown in Table \ref{timeDrone}. In this case, it is not possible to determine a 100\% probability of detection using drones because several factors, like whether the user has battery in the smartphone, the altitude of flight of the drone, the existence of obstacles between victim and drone, etc., hardly influence the results. However, as we saw in the previous section, if the flight altitude is not too high, the detection probability could increase, reaching at least a 90\%. However, if we consider that a grid can be created where the distance between the grid points is 50 m and the detection distance for each drone is at least 50 m, we have that when a drone is performing its route, each searching point of its area can intersect with the adjacent points searching areas. This means that each point could be visited from one to five times.

\begin{table}[htbp]
	\centering
	\caption{Needs to cover a 2.5 km squared area using a SAR team}
	\label{timeGeneralitat}
	\begin{tabular}{|l|l|l|l|r|}
		\hline
		Distance between people (m) & Nº people & Time each person (min.) & Total time (min.) & \multicolumn{1}{l|}{Pd} \\ \hline
		30       & 35   & 210  & 11.130 & 50\%          \\ \hline
		18       & 88   & 210  & 18.480 & 70\%          \\ \hline
		6       & 264  & 210  & 55.440 & 90\%          \\ \hline
	\end{tabular}
\end{table}

\begin{table}[htbp]
	\centering
	\caption{Needs to cover a 2.5 km squared area using drones}
	\label{timeDrone}
	\begin{tabular}{|l|l|l|l|c|}
		\hline
		Speed m/s & Nº Drones & Time each drone (min.) & Total time (min.) \\ \hline
		5  & 1   & 8.3    & 8.3           \\ \hline
		5  & 2   & 4.1    & 8.3           \\ \hline
		5  & 5   & 1.7    & 8.3           \\ \hline
	\end{tabular}
\end{table}

%----------------------------------------------------

\subsection{Alternatives to BLE Signal}

According to \cite{goursaud2015dedicated}, current wireless communications networks of the WAN type do not meet the requirements of the IoT elements such as communication range, costs and number of paired devices. Due to the possible deficiencies of the system seen in the previous subsections, the use of another type of technology is proposed that can deal with situations where the BLE signal is not good enough for an optimal detection of individuals. This is why new ones have been devised for the use of IoT devices such as LPWAN (Low-Power Wide Area Network), a type of connection similar to Wi-Fi networks that transmit a wireless signal lower than 1Ghz (compared to 2.4Ghz or 5Ghz of WiFi) that allows a greater signal wave and therefore, a greater range of communications. There are currently several different manufacturers, although the only one that has an open standard license is LoRa through LoRa-Alliance \citep{lora}. Another emerging communication technology, based on LPWAN, is Narrowband Fidelity (NB-Fi) defined in \citep{waviot}. This innovative technology is designed for communication of IoT devices focusing on the range and power consumption. The benefits of this technology include the type of free license, the connection range (16km in city and 50km in open country), security and scalability.

However, due to the specifications of the project presented in this paper, it is not possible to implement any of these field broadcast technologies because current smartphone devices do not have the integrated technology to receive and communicate with LPWAN. In a not far away future, when the devices contemplate the use of these technologies, it will be a feasible solution that will greatly improve the current one.

%--------------------------------------------------

\subsection{Cost per unit}

If a single rescue mission (or more than one) in a month requires several professional firefighters, the cost of the service increases proportionally with respect to the number of firefighters that are required. However, although the cost of a professional drone can be greater than the monthly salary of a firefighter, as long as the drones are not lost or damaged, the cost of the service will remain the same, independently of the number of missions. Thus, if we change an entire rescue team for a drone squadron, the initial cost may exceed the salary of the fire brigade, but when the time goes by the costs will be much lower than having an entire firefighter squad.

\section{Conclusions}
\label{sec:conclusions}

This paper presents a novel and practical system to find missing people using technologies like BLE beacons and UAVs. 
Drones are used to fly over an area in order to scan the BLE signal that is emitted by the missing person's smartphone. In this way, the proposed system can be used to improve SAR operations in emergency situations because it allows the emergency team to scan an area in less time using drones instead of the typical human emergency squad. Apart from the drones, there is no need to use any device other than smartphones. The coordination of all the parts that compose this proposal is performed using a web platform by an operator to create and establish a route before starting the mission. The generated data are processed in a back-end server that traces the route to make it optimal.
As a clear advantage of this proposal over other search systems, it actively helps emergency services and drastically reduces the time and cost of any operation. 
As future work, an updated version will be developed in a real environment in order to check its performance. In addition, a new device could be used to complement the task of drones in night flights, which will would the detection of people by patterns in images using a thermal camera.

\section*{Acknowledgements}
Research supported by the Spanish Ministry of Economy and Competitiveness, the FEDER Fund, and the CajaCanarias Foundation, under Projects TEC2014-54110-R, RTC-2014-1648-8, MTM2015-69138-REDT and DIG02-INSITU.

%
% ---- Bibliography ----
%
\bibliography{references.bib}
%%\bibliography{UAV-BLE-SAR2}
\bibliographystyle{spbasic}

%\bibliography{UAV-BLE-SAR}

\end{document}